\documentclass{iaus}
\usepackage{graphicx}

\title[The IMF in Extreme Star-Forming Environments] 
{The IMF in Extreme Star-Forming Environments: \\Searching for Variations vs. Initial Conditions}

\author[Andersen et. al]   
{Morten Andersen$^1$, M. R. Meyer$^1$, J. Greissl$^1$, B. D. Oppenheimer$^1$, M. A. Kenworthy$^1$, D. W., McCarthy$^1$, H. Zinnecker$^2$}

\affiliation{$^1$Steward Observatory, The University of Arizona, \\933 N. Cherry Avenue, Tucson, AZ 85721-0065\\ email: mandersen@as.arizona.edu, mmeyer@as.arizona.edu\\
$^2$Astrophysical Institute Potsdam, An der Sternwarte 16 14482 Potsdam, Germany}

\pubyear{2005}
\volume{IAUS227}  
\pagerange{119--126}
\date{?? and in revised form ??}
\jname{Proceedings Title IAU Symposium}
\editors{A.C. Editor, B.D. Editor \& C.E. Editor, eds.}
\begin{document}

\maketitle

\begin{abstract}
Any predictive theory of star formation must explain
observed variations (or lack thereof) in the initial 
mass function. 
Recent work suggests that we might 
expect quantitative variations in the IMF as a 
function of metallicity \cite[(Larson 2005)]{larson} or  magnetic field strength
\cite[(Shu et al.  2004)]{shu}. 
We summarize results from several on-going studies
attempting to constrain the ratio of high to low
mass stars, as well as stars to sub- stellar objects,
in a variety of different environments, all containing
high mass stars.

First, we examine the ratio of stars to sub--stellar
objects in the nearby Mon R2 region  utilizing 
NICMOS/HST data.  
We compare our results to the IMF by \cite[Kroupa (2002)]{kroupa2002} and to the observed ratios for IC 348 and Orion. 
Second, we present preliminary results
for  the ratio of high to low mass stars in W51, 
the most luminous {\mbox{H\,{\footnotesize II}}} region in the galaxy.   Based on 
ground--based multi--colour images of the cluster obtained
with the MMT adaptive optics system, we derive a lower limit to the ratio of high-mass to low-mass stars and compare it to the  ratios for nearby clusters. 
Finally,   we present the derived 
IMF for the R136 region in the LMC where the metallicity
is $1/4$ solar using HST/NICMOS data.  We find that the IMF is consistent with 
that characterizing the field \cite[(Chabrier 2003)]{chabrier}, as well 
as nearby star--forming regions, down to 1.0 M$_\odot$ outside 2 pc.  
Whereas the results for both Mon R2 and R136 are consistent with the nearby clusters, the ratio of high to low mass stars in W51 tentatively indicates a lack of low--mass objects.

\end{abstract}

\firstsection 
\section{Introduction}
What are the physical processes determining the shape of the IMF? 
Closely linked to this question is whether the IMF is universal or not. 
Studies of nearby, resolved, young star clusters indicate the IMF is remarkably similar,  $dN/ m\propto M^{-2.35}$ for  higher mass stars  and a flattening  around  0.5 M$_\odot$  where the slope changes from the Salpeter value to $\sim -1.35$ \cite[(Lada \& Lada 2003]{ladalada}. 
Although these studied indicate the stellar IMF might be roughly similar in nearby star forming regions above $\sim$ 0.1 M$_\odot$ \cite[(Meyer et al., 2000)]{meyer}, all the clusters studied to date essentially cover the same parameter space in terms of metallicity and cluster mass (to within an order of magnitude). 
In  the brown dwarf regime, there are indications  Taurus and Orion differ in the sense that Taurus is deficient in brown dwarfs relative to Orion \cite[(Luhman 2004)]{luhman}. 

For resolved stellar populations in more extreme environments, the evidence for variations in the IMF is relatively sparse. 
In the Arches cluster near the Galactic centre, the derived  IMF is flatter than Salpeter \cite[(Figer et al. 1999; Stolte et al. 2002)]{figer;stolte}. 
However, \cite[Portegies Zwart et al. (2002)]{zwart} showed  the flattening observed could be due to dynamical mass segregation. 
For the metal--poor  massive cluster R136 in the LMC, \cite[Sirianni et al. (2000)]{sirianni} presented an IMF that showed a distinct flattening at 2 M$_\odot$ down to 1.4 M$_\odot$. 

Theoretical considerations indicate the IMF should depend on environment. 
In particular it has been suggested it should depend on  the magnetic field \cite[(Shu et al. 2004)]{shu}  or on a critical density where the gas changes from a phase where it cools through molecular lines to a phase where it is heated due to  coupling to the dust  \cite[(Larson 2005)]{larson}. 
\cite[Shu et al. (2004)]{shu} suggested that the stellar mass is decided by the supercritical fraction of the molecular core which is inversely proportional to the magnetic field strength. 
To date, very little has been done in  searching for  variations in the IMF as a function of  mass to magnetic flux density ratio in star forming regions. 
The magnetic field component along the line of sight has been measured in some star forming molecular clouds  \cite[(Crutcher 1999)]{crutcher}, and  variations  in the mass to magnetic flux density ratio of a factor of 6 has been found. 
In contrast, \cite[Larson (2005)]{larson} has suggested a characteristic mass is associated with the critical density where the  gas undergoes a change in the polytropic index from  $\sim 0.7$ to $\sim 1.03$. 
Simulations have shown that an increase in the polytropic index tends to increase the average fragmentation mass  \cite[(Li et al. 2003)]{li}.  
The increase in the typical mass of a fragment results in a natural characteristic mass corresponding to the Jeans mass at the critical density. 
Since it is the interplay between cooling and heating that determines the critical density, it should be expected to vary with metallicity. 
Due to the lower amount of dust at low metallicities, the coupling of gas and dust occurs at a higher density and the Jeans mass at this point might actually be lower than for solar metallicity  \cite[(Larson 2005)]{larson}. 
Clusters with different metallicities can be found within the Galaxy by probing both inner and outer Galaxy young clusters.

Possibly the IMF is already imprinted in the molecular clump mass spectrum before any star formation has taken place.  
Combining the clump mass spectrum recently derived for IRAS19410+2336  for the mass range 3--20 M$_\odot$ \cite[(Beuther \& Schilke 2003)]{beuther} with the clump mass spectrum for the low mass region Rho Oph \cite[(Motte et al. 1998)]{motte}, the shape is similar to the stellar IMF. 
This could indicate that the stellar IMF might  not be determined by e.g. outflows and stellar winds. 
Probing clusters of different masses might indicate if this is true or not. 
Massive clusters have more massive stars and can therefore have a stronger effect on their environment through their UV flux and strong winds. 

Variations of the characteristic mass will have a strong effect on the relative numbers  of high- to low-mass stars. 
Changing the mass where the IMF flattens from a Salpeter power law to +1 dex shallower from  0.5 M$_\odot$ to 0.3 M$_\odot$ will lower the ratio of 1.0--10 M$_\odot$ stars to 0.1--1.0 M$_\odot$ by 30\%. 
A change to 0.7 M$_\odot$ will increase this ratio by 40\%. 
We have begun a study of possible variations in the IMF  by choosing clusters at different masses, metallicities and mass to magnetic flux density ratios. 
Here, we present  preliminary results for three of them: Mon R2, W51, and R136. 

\section{Mon R2, constraining the sub--stellar IMF}
With its earliest--type member  a B0 star, Mon R2 is expected to have a total mass intermediate between  IC 348 and Orion. 
Due to its relative youth ($\sim$ 1 Myr) and its proximity of 830 pc \cite[(Herbst et al. 1976)]{herbst}, the embedded cluster can be probed far below the hydrogen burning limit. 
We have obtained  near-infrared imaging data with HST/NICMOS 2 in order to  constrain the ratio of low mass stars  objects (0.1--1 M$_\odot$) to sub--stellar objects (0.04--0.1 M$_\odot$). 
A $3\times3$ mosaic covering 1\hbox{$^\prime$} (corresponding to 0.24$\times$0.24 pc at a distance of 830 pc)  was observed in the F110W, F160W, and the F220M filters, roughly corresponding to the J, H, and K filters.  
The data were transformed to the CIT system for comparison with the colors of normal stars and the interstellar reddening vector. 

\begin{figure}
 \includegraphics[width=12.5cm]{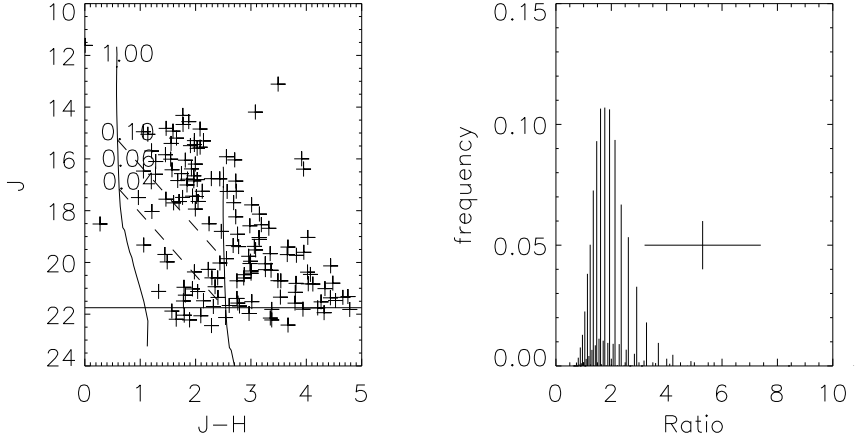}
  \caption{Left: the J--H, J colour--magnitude diagram for Mon R2. 
The solid lines are the 1 Myr  \cite[Baraffe et al. (1998)]{baraffe} isochrone projected to the distance of Mon R2 unreddened and reddened by A$_\mathrm{V}=17.5$ mag.  
The dashed lines illustrate the location of  the low--mass objects (0.04--0.1 M$_\odot$). 
Right: The observed ratio of the number of stars in the mass interval 0.1--1.0 M$_\odot$ over the number from  0.04--0.1 M$_\odot$, shown as a cross where the horizontal line marks the Poisson error. 
The histogram shows the probability distribution for obtaining a given ratio of low mass stars to sub--stellar objects when drawn from a \cite[Kroupa (2002)]{kroupa} single star IMF. }
\label{monR2_CMD}
\end{figure}

In total, we detect 182 sources brighter than $\mathrm{J_{cit}}=23.3$, showing a wide variety of extinction, as expected for a young embedded cluster. 
The J--H versus J colour--magnitude diagram is shown in Fig.~\ref{monR2_CMD} together with a 1 Myr \cite[Baraffe et al. (1998)]{baraffe} isochrone  and the same isochrone reddened by A$_\mathrm{V}=17.5$ mag. 
Following \cite[Wilking et al. (2004)]{wilking}, we de--redden to the 1 Myr isochrone for stars with no detected near-infrared excess and to the T-Tauri locus for stars with near-infrared excess \cite[(Meyer et al. 1997)]{meyer97}. 
In this way,  we have created an extinction limited sample between A$_\mathrm{V}=1$ mag and A$_\mathrm{V}=17.5$ mag. 
Stars with a de--reddened H--K colour $>$ 1.0 have been excluded because it is not clear how to interpret their colors.  
A total of   62 objects are located within the A$_\mathrm{V}$ limit imposed by the  completeness limit corresponding to  40 M$_\mathrm{jup}$, assuming an age of 1 Myr.

Because  the uncertainties on any individual estimate of  stellar mass are large we have chosen not to analyze the mass spectrum. 
Instead, we  derive the ratio of stars between 0.1 and 1 M$_\odot$ and objects in the mass range 0.04 M$_\odot$ to 0.1 M$_\odot$. 
We have compared that ratio with the predicted ratio from a \cite[Kroupa (2002)]{kroupa} single star IMF where the expected distribution of ratios is found by Monte Carlo experiments. 
Our derived ratio is  5.3 $\pm$ 2.1, significantly higher than the mode  found for the Monte Carlo experiments. 
Since we have not corrected for field stars, the ratio is expected to be a lower limit since field stars will be more numerous at faint magnitudes. 
However, due to the rather small field of view and since  Mon R2 is out of the Galactic plane we expect a small number of interlopers. 

The comparison with the single star IMF is problematic since that assumes a binary fraction of 0\%. 
Adopting the other extreme, all stars are in binaries, \cite[Kroupa (2001)]{kroupa2001} finds a much lower fraction of low--mass stars and brown dwarfs than the single star IMF. 
Thus, the  disagreement between the derived ratio and the predicted value  by the \cite[Kroupa 2002]{kroupa} IMF can be resolved  by assuming many of the stars in Mon R2 are in binaries. 
We will explore this further in Andersen et al., in prep. 

A similar ration can be derived for other regions. 
The same ratio as derived above  for IC 348 is 4.0$\pm$0.8, using the data from  \cite[Luhmann (2003)]{luhmanIC348}. 
Similarly, spectroscopic data for the Orion Nebula Cluster by \cite[Slesnick et al. (2004)]{slesnick} gives a ratio of 7.2$\pm1.6$. 
 Within the errors, the ratios derived for these two clusters seem  to be in agreement with  our findings here. 
More details can be found in \cite[Andersen et al., in prep. ]{andersen1}

\section{W51, a massive young cluster}
\begin{figure}
 \includegraphics[width=12.5cm]{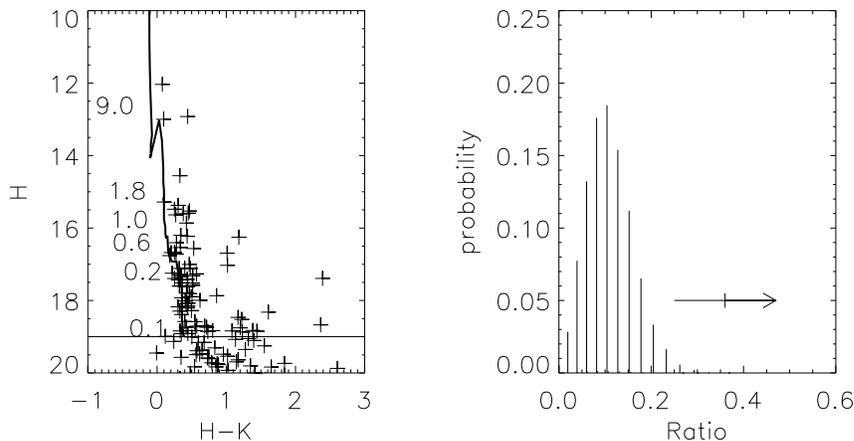}
  \caption{Left: the H--K, H colour--magnitude diagram for W51. 
The bold line is a \cite[Siess et al. (2002)]{siess} isochrone projected to the distance of W51, while the horizontal line us our estimated completeness limit. 
Right: The ratio  of stars in the mass interval 1.--1.0 M$_\odot$ over the stars from  0.1--1.0 M$_\odot$, shown as a cross where the horizontal line marks the Poisson error. 
The histogram shows the probability distribution for obtaining a given ratio of high--mass to low--mass objects when drawn from a \cite[Kroupa (2002)]{kroupa} single star IMF. }
\label{W51_CMD}
\end{figure}

W51 is the most  massive {\mbox{H\,{\footnotesize II}}} region in the Galaxy and is an excellent place  to determine the IMF in an extreme environment. 
At a distance of $\sim$ 7 kpc, deep near--infrared imaging can penetrate the dust in the young region and probe the IMF to sub--stellar masses. 
We have performed a preliminary study of a small region (30\hbox{$^{\prime\prime}$}$\times$30\hbox{$^{\prime\prime}$}, 0.1 pc$\times$0.1 pc)  in W51 using the MMT/ARIES Adaptive Optics  system \cite[(Kenworthy et al. 2004)]{kenworthy}. 
Total integration times were 600 seconds  in H and 1100 seconds in K and the effective seeing  for the images was 0\hbox{$.\!\!^{\prime\prime}$}14, enabling us to reach 0.1 M$_\odot$ through A$_\mathrm{V}=10$ mag. 
The colour--magnitude diagram plotted together with a \cite[Siess et al. (2000)]{siess} 1 Myr isochrone is shown  in Fig.~\ref{W51_CMD}. 
We de--reddened the sources to the presented isochrone by sliding the objects along the reddening vector, and thereby ignoring any potential near--infrared excess.    
Due to the uncertainties in the individual mass estimates, we only constrain the ratio of stars in a given mass range. 
The  data allows us to  determine the ratio of stars in the interval 1.0--10 M$_\odot$ to that from  0.1--1.0 M$_\odot$, where we find a ratio of 0.36$\pm$0.11. 
A comparison with the \cite[Kroupa (2002)]{kroupa} IMF is shown on the right side of Fig.~\ref{W51_CMD}.
Again,we find a ratio  higher than what is predicted by the \cite[Kroupa (2002)]{kroupa} IMF. 
Only three percent of the Monte Carlo simulations produce a ratio as high as we have observed within one sigma errors. 
The  lack of field star subtraction will make this value a lower limit as we expect significant contamination at the low mass end in this field. 

\cite[Meyer et al. (2000)]{meyer2000} present similar ratios for several nearby star--forming regions. 
Our derived value is significantly higher than  some of the regions, including the ONC and OMC--2, and Rho Oph  with ratios of 0.07$\pm$0.02,  0.07$\pm$0.04, and 0.1$\pm$0.04,  respectively. 
Within the errors, our measured ratio is consistent with   NGC 2024, Mon R2, RCrA and IC 348. 
A larger area sampled with longer integrations times together with a control field will place much better constraints on this ratio.  
We should also  be able to sample to lower masses, beyond the hydrogen burning limit.

\section{The low--mass stars in R136}
Powering the most luminous {\mbox{H\,{\footnotesize II}}} region in the Local Group, 30 Dor is considered a scaled-down version of  distant star bursts. 
At a distance of 50 kpc, it is possible to resolve the stellar population down to $\approx$ 1 M$_\odot$ and thus derive the IMF by direct star counts. 
Using HST/WFPC2 observations, \cite[Sirianni et al. (2000)]{sirianni} derived the IMF down to 1.35 solar masses and found a distinct flattening of the IMF at 2 M$_\odot$ at a distance of $\sim$ 1.5 pc from the center.  
However, the authors did not take into account that R136 suffers from differential extinction, A$_\mathrm{V}$=1--3 mag \cite[(Brandl et al. 1996)]{brandl}. 

\begin{figure}
 \includegraphics[width=12.5cm]{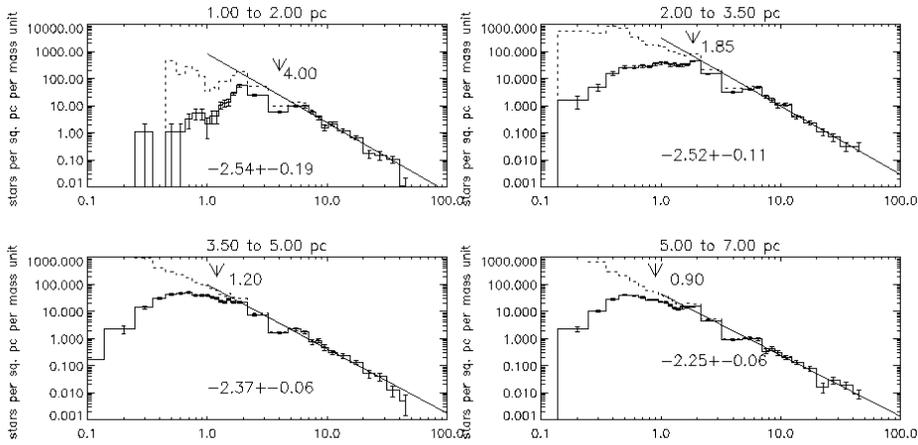}
  \caption{The derived IMFs for four different annuli around R136. 
The solid histograms are the derived IMF without taking incompleteness into account; the dashed histograms are the IMFs corrected for incompleteness. }
\label{R136_res}
\end{figure}

We have observed the central one arcminute of the 30 Dor cluster, which has R136 in its centre, using  HST/NICMOS 2 with the F160W filter. 
The total integration time per position in the mosaic is 1 hour. 
Source detection was done with great care to avoid identifying diffraction features as stars and in total, photometry was obtained for some 10000 stars within our field of view. 
Fig.~\ref{R136_res} shows the derived IMF outside 1 pc, assuming an age of 3 Myr,  justified by the presence of WR stars and the location of  pre--main sequence stars in the colour--magnitude diagrams.  
Outside 1 pc, the slopes are consistent with a Salpeter slope. 
In particular, there is no evidence for a flattening at 2 M$_\odot$ as found in the optical HST data set by \cite[Sirianni et al.]{sirianni} 
We believe the different results can be explained  of differential extinction within the cluster which was not taken into account by \cite[Sirianni et al.]{sirianni} 
By not creating an extinction limited sample, preferentially  the lower mass stars will not be detected. 
More details can be found in \cite[Andersen et al. \textit{in prep}]{andersen2}
\section{Conclusions}
We have presented results on the IMF for the three young star forming regions containing massive stars: R136, W51, and Mon R2. 
These three clusters cover a large range of cluster masses from less than $10^3$ M$_\odot$ for Mon R2 to around $10^5$ M$_\odot$ for R136. 
Despite the lower metallicity, the  IMF in R136 appears to be a power law with a Salpeter slope down to  1 M$_\odot$.  
Similarly, the ratio of 0.1--1 M$_\odot$ stars over 0.04--0.1 M$_\odot$ in Mon R2 is consistent with the same ratio derived for IC 348 and Orion. 
On the other hand, we find a deficit of low-mass objects in W51 relative to nearby less massive clusters like ONC and OMC--2, although our derived ratio is consistent with several other nearby regions. 
Future observations of  a larger region in W51 will enable us to reduce the error bars on our derived ratio and hence firmly establish if the IMF in W51 is ``bottom light''.

%

\begin{acknowledgments}
We would like to thank our collaborators in the different projects: Bernhard Brandl, Wolfgang Brandner, John Carpenter, Angela Cotera, Catherine Dougados,  Lynne Hillenbrand, Georges Meylan, and  Andrea Moneti. 
MA was partially funded through DLR grant 50OR9912. 
This work was supported by a Cottrell Scholar's Award to MRM from the Research Corporation and NASA grant HST13-9846
\end{acknowledgments}

\end{document}